# Inertia tensor and cross product
# In n-dimensions space

M. Hage-Hassan
Université Libanaise, Faculté des Sciences Section (1)
Hadath-Beyrouth

**Abstract**

We demonstrated using an elementary method that the inertia tensor of a material point and the cross product of two vectors were only possible in a three or seven dimensional space. The representation matrix of the cross product in the seven dimensional space and its properties were given. The relationship between the inertia tensor and the octonions algebra was emphasized for the first time in this work.

**Résumé**
Nous montrons par une méthode élémentaire que le tenseur d'inertie d'un point matériel et le produit vectoriel de deux vecteurs sont possibles seulement si la dimension de l'espace est 3 ou 7. La représentation matricielle dans l'espace de 7 dimensions ainsi que ses propriétés sont données. La relation entre le tenseur d'inertie et l'algèbre des octonions est soulignée pour la première fois dans ce travail.

**1. Introduction**
The vector cross product in the Euclidean space of 3 dimensions is largely used in physics, but the generalization by Eckmann (1-2) to 7 dimensions is not well known by the physicists. This generalization starts to be useful in modern physics (2-3) and a simple presentation to make these concepts available is interesting. We present these concepts on the basis of the inertia tensor and its generalization (4). This allows us by a simple method to obtain and to present the concepts of quaternion and octonion as well as there representation matrix and its properties.

**2. Inertia Tensor**
The kinetic energy of a particle of mass m=1 which moves in a system in rotation with angular velocity $(\vec{\omega})$ is $T = \frac{1}{2}(\vec{\omega} \times \vec{r}) \cdot (\vec{\omega} \times \vec{r})$.

With, $\vec{X} = \vec{\omega} \times \vec{r}$ this is written in the matrix form $(X) = (V_3)(\omega)$

$$\begin{pmatrix} x_1 \\ x_2 \\ x_3 \end{pmatrix} = \begin{pmatrix} 0 & z & -y \\ -z & 0 & x \\ y & -x & 0 \end{pmatrix} \begin{pmatrix} \omega_1 \\ \omega_2 \\ \omega_3 \end{pmatrix} \qquad (1)$$

The kinetic energy became: $T = \frac{1}{2}(X)^t \cdot (X) = \frac{1}{2}(\omega)^t (V_3)^t (V_3)(\omega)$

$(X)^t$ is the transpose of (X).

We write the inertia matrix: $(M) = (V_3)^t (V_3)$

$$(M) = \begin{pmatrix} m_{11} & m_{12} & m_{13} \\ m_{12} & m_{22} & m_{23} \\ m_{13} & m_{23} & m_{33} \end{pmatrix} = \begin{pmatrix} \vec{r}^2 - x^2 & -xy & -xz \\ -xy & \vec{r}^2 - y^2 & -yz \\ -xz & -yz & \vec{r}^2 - z^2 \end{pmatrix} \quad (2)$$

## 3. Inertia tensor and the quaternions

The identification of two sides of the equation (2) may be written as:

$$m_{11} + x^2 = \vec{r}^2 \quad m_{12} + xy = 0 \quad m_{13} + xz = 0$$
$$m_{12} + xy = 0 \quad m_{22} + y^2 = \vec{r}^2 \quad m_{23} + yz = 0$$
$$m_{13} + xz = 0 \quad m_{23} + yz = 0 \quad m_{33} + z^2 = \vec{r}^2$$

with $\quad x^2 + y^2 + z^2 = \vec{r}^2$.

We can express these systems in matrix form as $(H_4)^t (H_4) = \vec{r}^2 I$

$$\begin{pmatrix} & & & -x \\ (V_3)^t & & & -y \\ & & & -z \\ x & y & z & 0 \end{pmatrix} \begin{pmatrix} & & & x \\ (V_3) & & & y \\ & & & z \\ -x & -y & -z & 0 \end{pmatrix} = \begin{pmatrix} \vec{r}^2 & 0 & 0 & 0 \\ 0 & \vec{r}^2 & 0 & 0 \\ 0 & 0 & \vec{r}^2 & 0 \\ 0 & 0 & 0 & \vec{r}^2 \end{pmatrix} \quad (3)$$

$I$ is the identity matrix.

We replace the matrix $(V_3)$ by its expression in (1), we deduce the orthogonal and antisymmetric matrix:

$$(H_4) = \begin{pmatrix} 0 & z & -y & x \\ -z & 0 & x & y \\ y & -x & 0 & z \\ -x & -y & -z & 0 \end{pmatrix} \quad (4)$$

The matrix $(H_4)$ is the matrix representation of the quaternion $h = xe_1 - ye_2 + ze_3$

With
$$e_1^2 = -1, \quad e_2^2 = -1, \quad e_3^2 = -1$$
$$e_1 e_2 = e_3, \quad e_2 e_3 = e_1, \quad e_3 e_1 = e_2$$
(5)

$(H_4)$ Is the Hurwitz matrix and $e_1, e_2$ and $e_3$ are the generators of the algebra.

## 4. Inertia tensor and the octonions

If we write $\vec{r} = \sum_{i=1}^{n} x_i \vec{e}_i$, with $\{\vec{e}_i\}$ is the base of the vector in Euclidean space $R^n$. The generalization of the tensor of inertia in an intuitive way (4) is written then:

$$(M) = \begin{pmatrix} m_{11} & m_{12} & \ldots & m_{1n} \\ m_{21} & m_{22} & \ldots & m_{2n} \\ \vdots & \ldots & \ldots & \vdots \\ m_{1n} & m_{2n} & \ldots & m_{nn} \end{pmatrix} = \begin{pmatrix} \vec{r}^2 - x_1^2 & -x_1 x_2 & \ldots & -x_1 x_n \\ -x_1 x_2 & \vec{r}^2 - x_2^2 & \ldots & -x_2 x_n \\ \vdots & \ldots & \ldots & \vdots \\ -x_1 x_n & -x_2 x_n & \ldots & \vec{r}^2 - x_n^2 \end{pmatrix} \quad (6)$$

The identification of two members gives:
$$m_{ii} + x_i^2 = \vec{r}^2, i = 1,\ldots,n$$
$$m_{ij} + x_i x_j = 0, i \neq j \text{ et } i,j = 1,\ldots,n$$

And the matrix system $(H_n)^t (H_n) = \vec{r}^2 I$ takes the form:

$$\begin{pmatrix} & -x_1 & \\ (V_n)^t & \vdots & \\ & -x_n & \\ x_1 & \ldots & x_n & 0 \end{pmatrix} \begin{pmatrix} & x_1 & \\ (V_n) & \vdots & \\ & x_n & \\ -x_1 & \ldots & -x_n & 0 \end{pmatrix} = \begin{pmatrix} \vec{r}^2 & 0 & \ldots & 0 \\ 0 & \vec{r}^2 & \ldots & 0 \\ \vdots & \ldots & \ldots & \vdots \\ 0 & \ldots & \ldots & \vec{r}^2 \end{pmatrix} \quad (7)$$

Hurwitz (5) showed that we can only build orthogonal and antisymmetric matrix which lines are a linear combination of components of a vector only if n=1, 2, 4 or 8. Consequently the matrix $(H_n)$ or Hurwitz matrix is orthogonal if n+1=8, it results from it that dim ($R^n$) = 1, 3 or 7.

The matrix $(H_8)$ is the matrix representation of the octonion $h = \sum_{i=1}^{i=7} x_i e_i$.
The generators of the algebra satisfy:
$$e_i^2 = -1, \quad i = 1,\ldots 7$$
$$e_i e_j = -e_j e_i \quad (8)$$

We can obtain this matrix by hand easily, which will be the object of paragraph 6.

## 5. Dimension of $R^n$ and cross product

The problem of the research of the dimension of space where we define the cross product is known for a long time. We know that dim=3 or 7 and we will simply find all these results starting from number of parameters of $(V_n)$ which is $\frac{n(n-1)}{2}$. Moreover lines of the matrix (V) are made of components of vector $\vec{r} = \sum_{i=1}^{n} x_i \vec{e}$.

1- In the case where the vector is without component
$\frac{n(n-1)}{2} = 0$ the solution is: n=0 and n=1 therefore dim ($R^n$) =1.

2- In the case where the vector have n components
$\frac{n(n-1)}{2} = n$ the solution is: n=0 and n=3 therefore dim ($R^n$) =3.

3- In the case where n>3 the vector has n components that are subjected to the constraints conditions:
$$(\vec{e}_i \times \vec{r}).\vec{e}_i = (\vec{e}_i \times \vec{r}) \cdot \vec{r} = 0,$$

The number of the constraints is 2n then it results from it that
$\frac{n(n-1)}{2} = n + 2n$ the solution is: n=0 and n=7 therefore dim ($R^n$) =7.

We thus checked in a simple way the theorem of Eckmann (6-7).

## 6. Hurwitz's Transformation and it's matrix representation.

To determine the matrices (H) we must notice that these matrices are antisymmetric and orthogonal. Moreover if $\vec{\omega} = \vec{r} = \vec{u}$ we find the relation on sums of squares (5), that we write

$$(Z)^t (Z) = (\vec{u}^2)^2$$

with
$$\vec{Z} = (z_1, \ldots, z_n) \ et \ \vec{u}^2 = u_1^2 + \ldots + u_N^2.$$

In what follows, we will expose by two simple methods of recurrences (8) the determination of the matrices (V). The first one takes its starting point the transformation of Levi-Civita and the orthogonality of the matrices (H), the second is based on the law of composition algebra of Cayley-Dickson.

### 6.1 Levi-Civita Transformation

For n=2 Levi-Civita introduced the conformal transformation which is an application of $R^2 \rightarrow R^2$.

$$z_1 = u_1^2 - u_2^2, \quad z_2 = 2u_1 u_2$$

that is written
$$\begin{pmatrix} z_1 \\ -z_2 \end{pmatrix} = \begin{pmatrix} u_1 & u_2 \\ -u_2 & u_1 \end{pmatrix} \begin{pmatrix} u_1 \\ -u_2 \end{pmatrix} = (H_2)(U_2) \qquad (9)$$

### 6.2 Hurwitz's Transformations

For the generalization of the transformations of Levi-civita we pose

$$\begin{pmatrix} z_1 \\ z_2 \end{pmatrix} = 2 \begin{pmatrix} u_1 & u_2 \\ -u_2 & u_1 \end{pmatrix} \begin{pmatrix} u_3 \\ u_4 \end{pmatrix} = 2(H_2)(U_2')$$

Using the orthogonality of $(H_2)$ we find

$$z_1^2 + z_2^2 = 2(u_1^2 + u_2^2)(u_3^2 + u_4^2)$$

And if we put
$$z_3 = (u_1^2 + u_2^2) - (u_3^2 + u_4^2)$$

We write

$$\begin{pmatrix} z_1 \\ -z_2 \\ -z_3 \\ 0 \end{pmatrix} = \begin{pmatrix} u_3 & u_4 & u_1 & u_2 \\ -u_4 & u_3 & u_2 & -u_1 \\ -u_1 & -u_2 & u_3 & u_4 \\ -u_2 & u_1 & -u_4 & u_3 \end{pmatrix} \begin{pmatrix} u_1 \\ u_2 \\ u_3 \\ u_4 \end{pmatrix} = (H_4)(U_4) \qquad (10)$$

Thus we find the transformation of $R^4 \rightarrow R^3$ known by the transformation of Kustaanheimo-steifel.

To obtain $(H_8)$ and $(H_{16})$ we repeat the same process while replacing $(H_2), (U_2')$ and $z_3$ by $(H_4), (U'_4)^t = (u_5, \ldots, u_8)$ and $z_5$ we deduce $(H_8)$ then we adopt the same way for $(H_{16})$.

### 6.3 Hurwitz's Transformations and Cayley-Dickson algebra

We determine the matrix $(H_8)$ using the result of our work (9) on the transformation of Hurwitz in the theory of the angular momentum and which consists in posing

$$\begin{pmatrix} x & y \\ -\overline{y} & \overline{x} \end{pmatrix} = 2 \begin{pmatrix} \overline{v}_1 & -v_2 \\ \overline{v}_2 & v_1 \end{pmatrix} \begin{pmatrix} v_3 & v_4 \\ -\overline{v}_4 & \overline{v}_3 \end{pmatrix} \qquad (11)$$

With
$$x = z_1 + iz_2, \quad y = z_3 + iz_4,$$
$$v_1 = u_1 + iu_2, \quad v_2 = u_3 + iu_4$$
$$v_3 = u_5 + iu_6, \quad v_4 = u_7 + iu_8 \qquad (12)$$

We obtain
$$z_5 = (v_1 \overline{v}_1 + v_2 \overline{v}_2) - (v_3 \overline{v}_3 + v_4 \overline{v}_4)$$
$$z_1 + iz_2 = 2(\overline{v}_1 v_3 + v_2 \overline{v}_4), \qquad z_3 + iz_4 = 2(\overline{v}_1 v_4 - v_2 \overline{v}_3),$$
$$z_5 = |v_1|^2 + |v_2|^2 - |v_3|^2 - |v_4|^2.$$

After identification and a rather simple arrangement we obtain:

$$\begin{pmatrix} z_1 \\ -z_2 \\ -z_3 \\ -z_4 \\ -z_5 \\ 0 \\ 0 \\ 0 \end{pmatrix} = \begin{pmatrix} u_5 & u_6 & u_7 & u_8 & u_1 & u_2 & u_3 & u_4 \\ -u_6 & u_5 & u_8 & -u_7 & u_2 & -u_1 & -u_4 & u_3 \\ -u_7 & -u_8 & u_5 & u_6 & u_3 & u_4 & -u_1 & -u_2 \\ -u_8 & u_7 & -u_6 & u_5 & u_4 & -u_3 & u_2 & -u_1 \\ -u_1 & -u_2 & u & -u_4 & u_5 & u_6 & u_7 & u_8 \\ -u_2 & u_1 & -u_4 & u_3 & -u_6 & u_5 & -u_8 & u_7 \\ -u_3 & u_4 & u_1 & -u_2 & -u_7 & u_8 & u_5 & -u_6 \\ -u_4 & -u_3 & u_2 & u_1 & -u_8 & -u_7 & u_6 & u_5 \end{pmatrix} \begin{pmatrix} u_1 \\ u_2 \\ u_3 \\ u_4 \\ u_5 \\ u_6 \\ u_7 \\ u_8 \end{pmatrix} \qquad (13)$$

Finally if we write $\vec{r} = \sum_{i=1}^{n} x_i \vec{e}_i$ we deduce the matrix $(V_7)$:

$$(V_7) = \begin{pmatrix} 0 & x_7 & -x_6 & -x_5 & x_4 & x_3 & -x_2 \\ -x_7 & 0 & -x_5 & x_6 & x_3 & -x_4 & x_1 \\ x_6 & x_5 & 0 & x_7 & -x_2 & -x_1 & -x_4 \\ x_5 & -x_6 & -x_7 & 0 & -x_1 & x_2 & x_3 \\ -x_4 & -x_3 & x_2 & x_1 & 0 & x_7 & -x_6 \\ -x_3 & x_4 & x_1 & -x_2 & -x_7 & 0 & x_5 \\ x_2 & -x_1 & x_4 & -x_3 & x_6 & -x_5 & 0 \end{pmatrix} \qquad (14)$$

To determine the matrices $(H_{2^n}), n = 4, 5, \ldots$, we suppose that elements x, y, $v_1, v_2, v_3$ and $v_4$ are defined on H, O and more generally they may be elements of the algebra of Cayley-Dickson (9). Then we adopt the same method of recurrence as above.

### 7. Properties of the matrix (V)

In the case where n=7 we find for the matrix $(V_7)$ analogue properties of the matrix $(V_3)$ as follows:

$$(V_3)^3 = -(\vec{r}^2)(V_3)$$
$$(V_7)^3 = -(\vec{r}^2)(V_7) \qquad (15)$$

If the vector is unitary the following expression is valid for n=3 and n=7.
$$Exp[-i\theta(V_n)] = 1 - i\sin\theta(V_n) - (1 - \cos\theta)(V_n)^2 \qquad (16)$$

Consequently we find a striking analogy between the two cases:
$$(H_4)^2 = -I, \quad (V_3)^3 = -(V_3)$$
$$(H_8)^2 = -I, \quad (V_7)^3 = -(V_7) \qquad (17)$$

## 8. Hurwitz transformation and spinor theory

There is a close link between the Hurwitz transformations and spinor theory. In this regard, we put $z_i$ in quadratic form in terms of $(v)^t$ and $(v)$

### 8.1 Transformation $R^8 \to R^2$.

$$z_i = (v)^t (\sigma_i)(v)$$

With $(v)^t = (v_1 v_2)$ and $(\sigma_i)$ denotes the Pauli matrices (11).

$$(\sigma_1) = \begin{pmatrix} 0 & 1 \\ 1 & 0 \end{pmatrix}, \quad (\sigma_2) = \begin{pmatrix} 0 & -i \\ i & 0 \end{pmatrix}, \quad (\sigma_3) = \begin{pmatrix} 1 & 0 \\ 0 & -1 \end{pmatrix}$$

### 8.2 Transformation $R^8 \to R^5$.

Put $(v)^t = (v_1 \ v_2 \ v_3 \ v_4)$, then by explicit calculation we find
$$z_1 = (v)^t \gamma^5 (v), \quad z_2 = i(v)^t \gamma^2 (v)$$
$$z_3 = i(v)^t \gamma^2 (v), \quad z_4 = i(v)^t \gamma^1 (v)$$
$$z_4 = (v)^t \gamma^0 (v)$$

It is clear that γ-matrices are the famous Dirac representation.

$$\gamma^0 = \begin{pmatrix} I & 0 \\ 0 & I \end{pmatrix}, \quad \gamma^i = \begin{pmatrix} 0 & \sigma_i \\ -\sigma_i & 0 \end{pmatrix}, \quad \gamma^5 = \begin{pmatrix} 0 & I \\ I & 0 \end{pmatrix}$$

Finally we can change the Euclidean by a pseudo-Euclidean space (11) which doesn't affect our treatment. Moreover the study of Hurwitz transformations will be the subject of another paper.